# Optoelectronic Properties of Carbon Nanorings: Excitonic Effects from Time-Dependent Density Functional Theory

*Bryan M. Wong\**

Materials Chemistry Department, Sandia National Laboratories, Livermore, California 94551

*Corresponding author. E-mail: bmwong@sandia.gov

The electronic structure and size-scaling of optoelectronic properties in cycloparaphenylene carbon nanorings are investigated using time-dependent density functional theory (TDDFT). The TDDFT calculations on these molecular nanostructures indicate that the lowest excitation energy surprisingly becomes larger as the carbon nanoring size is increased, in contradiction with typical quantum confinement effects. In order to understand their unusual electronic properties, I performed an extensive investigation of excitonic effects by analyzing electron-hole transition density matrices and exciton binding energies as a function of size in these nanoring systems. The transition density matrices allow a global view of electronic coherence during an electronic excitation, and the exciton binding energies give a quantitative measure of electron-hole interaction energies in the nanorings. Based on overall trends in exciton binding energies and their spatial delocalization, I find that excitonic effects play a vital role in understanding the unique photoinduced dynamics in these carbon nanoring systems.

## 1. Introduction

Cyclic organic structures comprise a wide variety of interesting examples in strained, distorted, bent, and sterically-hindered molecular systems in nanoscience. These structures have fascinated chemists for several decades not only for their unique chemical bonding but also because of the potential to tailor their unique physical properties through variations in molecular size, shape, and crystallinity.[1-5] Cyclic nanostructures composed entirely of conjugated carbon subunits are especially fascinating since they exhibit a myriad of interesting electronic properties. In particular, since cycloparaphenylene nanostructures are composed of aromatic subunits with radially-oriented $p$ orbitals, these materials can support mobile charge carriers on the delocalized valence and conduction orbitals along the backbone chain.[5-7] As a result, cycloparaphenylenes and other well-ordered macrocyles may find applications as conducting materials in molecular electronics, organic field-effect transistors (OFETs), and nonlinear optics (NLOs) due to their unique optoelectronic properties.[8-13] In general, the ability to control and understand the electronic response of organic nanostructures at the molecular level would have a tremendous impact in nanoscale optoelectronic technologies.[14,15]

Carbon nanorings large enough to encapsulate fullerenes were first synthesized by Kawase et al.[5], and are composed of a conjugated carbon skeleton in a smooth belt-shaped structure. Many of the nanorings synthesized by the Kawase group are actually part of a larger family of molecules known as phenylacetylene macrocycles (PAMs) which are composed of alternating phenyl and ethynyl subunits sequentially bonded to form a single ring. The phenyl subunits in PAMs are usually linked in the *meta* or *para* positions, and can be composed of four to nine phenylene-ethynylene units.[16] In a very recent communication, Jasti et al. synthesized and characterized a set of [9]-, [12]-, and [18]-cycloparaphenylenes composed solely of phenyl rings sequentially connected to form a single "nanohoop" (Figure 1).[17] Their synthesis was especially noteworthy since cycloparaphenylenes form the fundamental annular segments of armchair carbon nanotubes and may serve as templates toward the rational synthesis of carbon nanotubes with specific chiralities.[17-19]

A peculiar feature of these cycloparaphenylenes is their unusual optoelectronic properties as a function of molecular size. Specifically, as the carbon nanoring size is increased, the lowest computed absorption energy becomes larger. This result is surprising and counterintuitive to our usual expectations! For example, our chemical intuition from organic chemistry tells us that as the number of repeat units in a macromolecule increases, the distance between molecular energy levels diminishes, effectively decreasing the optical absorption gap. This trend (which actually arises from quantum confinement effects[20a]) qualitatively describes the variation of absorbance properties in $\pi$-conjugated systems as the number of monomer units increase in a polymer. However, in the case of cyclic nanorings, the excitation energies as a function of size seem to contradict this quantum-confinement-effect argument.[20b] To rationalize these trends, with a view of understanding the nature of electronic excitations in these nanorings, I carried out *ab initio* calculations on both cyclic and linear paraphenylenes (Figure 2) comprised of 5 to 18 phenyl repeat units each. Results obtained from time-dependent density functional theory (TDDFT) give electronic band gaps, optical absorption energies, and oscillator strengths which can be used to estimate exciton binding energies as a function of paraphenylene size. The TDDFT calculations also provide a two-dimensional real-space analysis of transition densities which represent coherent electronic transitions between ground- and electronically-excited states.[21-25] These transition densities give a panoramic view of electron-hole coherence and exciton delocalization in each of the paraphenylene systems. Based on overall trends in exciton binding energies and their spatial delocalization, I find that excitonic effects play a significant role in understanding the unique photoinduced dynamics of cycloparaphenylene nanoring systems.

## 2. Theory and Methodology

The cyclic and linear paraphenylenes ($N = 5 - 18$) analyzed in this work are shown in Figure 2. All quantum chemical calculations were carried out with the parameter-free PBE0 hybrid density functional which incorporates a fixed combination of 25% Hartree-Fock exchange and Perdew's GGA corrections in the correlation contribution.[26] Previous investigations by Tretiak et al. have shown that the use of pure local and gradient-corrected functionals (i.e., LDA and PBE) results in unphysical unbound exciton states, while hybrid functionals partially overcome this problem by mixing in a fraction of nonlocal Hartree-Fock exchange.[24,25] Based on their studies, I have chosen the PBE0 functional for this work since the PBE0 kernel provides a balanced description of neutral excitons (both singlets and triplets) in conjugated polymers and carbon nanotubes.[24,25,27-30]

Ground-state geometries of all cyclic and linear paraphenylenes were optimized at the PBE0/6-31G(d,p) level of theory. Geometry optimizations were calculated without symmetry constraints, and root-mean-squared forces converged to within 0.00003 a.u. At the optimized ground-state geometries, TDDFT calculations were performed with a larger, diffuse 6-31+G(d,p) basis set to obtain the lowest four singlet vertical excitations. For both the ground-state and TDDFT calculations, I used a high-accuracy Lebedev grid consisting of 96 radial and 302 angular points. All *ab initio* calculations were performed with a locally modified version of GAMESS.[31]

Within the TDDFT formalism, one obtains the excited-state electron density, $\rho(\mathbf{r}) \equiv |\psi(\mathbf{r})|^2 = \sum_{i=1}^{N} |\phi_i(\mathbf{r})|^2$ composed of $N$-occupied molecular orbitals $\phi_i(\mathbf{r})$ as solutions from the time-dependent Kohn-Sham equations. In the same way that one can calculate transition density matrices from time-dependent Hartree-Fock orbitals, one can use orbitals from the noninteracting Kohn-Sham system to form transition densities which include many-body correlation effects from the TDDFT formalism. Following the two-dimensional real-space analysis approach of Tretiak et al.[21-25], one can construct coordinate $\mathbf{Q_v}$ and momentum $\mathbf{P_v}$ matrices with elements given by

$$(Q_v)_{mn} = \left\langle \psi_v \left| c_m^\dagger c_n \right| \psi_g \right\rangle + \left\langle \psi_g \left| c_m^\dagger c_n \right| \psi_v \right\rangle \tag{1}$$

$$(P_v)_{mn} = \left\langle \psi_v \left| c_m^\dagger c_n \right| \psi_g \right\rangle - \left\langle \psi_g \left| c_m^\dagger c_n \right| \psi_v \right\rangle \tag{2}$$

where $\psi_g$ and $\psi_v$ are ground and excited states, respectively. The Fermi operators $c_i^\dagger$ and $c_i$ represent the creation and annihilation of an electron in the $i$th basis set orbital in $\psi$. For the cyclic and linear paraphenylenes analyzed in this work, the $\mathbf{Q_v}$ and $\mathbf{P_v}$ matrices each form a two-dimensional $xy$-grid running over all the carbon sites along the $x$- and $y$-axes. The specific ordering of the carbon sites used in this work is shown in Figure 2. The $(P_v)_{mn}$ momentum matrix represents the probability amplitude of an electron-hole pair oscillation between carbon sites $m$ and $n$, and the $(Q_v)_{mn}$ coordinate matrix gives a measure of exciton delocalization between sites $m$ and $n$, respectively. Each of these matrices provides a global view of electron-hole coherence and exciton delocalization for optical transitions within the paraphenylene systems.

## 3. Results and Discussion

### 3.1. Geometries and excitation energies

In the ground-state optimizations for the linear paraphenylenes, the dihedral angle between adjacent benzene rings was calculated to be 37°. This is close to the dihedral angle in isolated biphenyl, and is consistent in all of the linear paraphenylenes regardless of length. In contrast, for the smallest and highly-strained $N = 5$ cyclic paraphenylene, the benzene rings adopt a wide variety of smaller dihedral angles between 9° and 27°. Figure 3 shows the average dihedral angle between adjacent benzene rings as a function of paraphenylene size. For a given number of benzene rings, the average dihedral angle in the cyclic paraphenylenes is always smaller than their acyclic counterparts. However, as the size of the cyclic paraphenylene increases, the strain energy becomes smaller, and the dihedral angles between benzene rings increase to 36°. The optimized Cartesian coordinates for both the cyclic and linear paraphenylenes can be found in the Supplementary Information.

To investigate optoelectronic trends as a function of size and shape, optical absorption energies, $E_{opt}$, were computed for all the cyclic and linear ground-state geometries. Table 1 compares the lowest excitation energies and oscillator strengths between the cyclic and linear paraphenylenes, and Figure 4 displays $E_{opt}$ as a function of size. The other higher-lying singlet excitations (up to $S_4$) can be found in the Supplementary Information. As the number of benzene rings increases, the lowest excitation energies for both the cyclic and linear paraphenylenes asymptotically approach a value of 3.48 eV. However, as Figure 4 illustrates, the manner in which they approach this asymptotic value is considerably different: $E_{opt}$ for the cyclic systems increases with size, while $E_{opt}$ for the acyclic systems decreases. These unusual trends are discussed further in Section 3.3.

Another significant difference between the two systems is the variation of oscillator strengths as a function of size. For the linear paraphenylenes, the oscillator strengths increase linearly as a function of chain length, while the oscillator strengths for most of the even-membered cyclic systems are zero (due to molecular symmetry). Although the DFT geometry optimizations were performed without symmetry constraints, most of the calculations for the smaller ($N \leq 12$) even-membered nanorings converged to a $C_{2v}$-like symmetric structure. For nanorings containing an odd number of benzenes, the optimized structures have a reduced symmetry, and the oscillator strengths are nonzero. However, as the number of benzene rings increases, the strain energy becomes smaller, and several conformers with different dihedral angles can exist with similar energies. I found that these different conformers have various oscillator strengths (due to reduced symmetry) but nearly indentical energies. A complete analysis of the conformational landscape available to the nanorings is beyond the scope of this work and would have a negligible effect on the excitation energies in the larger systems. The small $S_1 \leftarrow S_0$ oscillator strengths obtained for the nanorings also have a close analogy with the very recent study by Kilina et al. which calculated optoelectronic properties of finite-length carbon nanotubes.[32] In their study, it was found that hybrid functionals determine the lowest singlet-excited state to be an optically inactive ("dark") state with the optically active ("bright") state lying higher in energy. Similarly, for the large ($N \geq 12$) nanorings in this work, the second and third singlet excitations are bright states with strongly-allowed transitions (Table SI-1 in the Supplementary Information). These trends in oscillator strengths for the nanorings can also be explained qualitatively in terms of a simple exciton model where one transition dipole moment is assigned to each benzene ring (Figure 5). For the lowest $S_1 \leftarrow S_0$ excitation, the transition moments in both the linear and cyclic paraphenylenes are aligned in a head-to-tail arrangement. However, because of their circular geometries, the $S_1 \leftarrow S_0$ transition moments in the cyclic paraphenylenes effectively cancel, while the total transition moment increases as a function of length in the linear systems (for the larger, more flexible nanorings, some of the transition dipole moments have a small component perpendicular to the ring, and the vectorial sum is slightly nonzero). In contrast, for the other higher-lying singlet excitations, the transition dipoles are aligned in one direction for half of the nanoring and in the opposite direction for the other half of the ring, producing a net transition dipole and a nonzero oscillator strength. As a result, the structural difference between cyclic and linear geometries imposes additional symmetry constraints which determine the oscillator strengths in these conjugated systems.

### 3.2. Transition Density Matrix Analysis

To provide further insight into these optoelectronic trends, I carried out a two-dimensional real-space analysis of density matrices for both the cyclic and linear systems. Figure 6 displays the absolute values of the coordinate density matrix elements, $\left|\left(Q_1\right)_{mn}\right|$, for the lowest excitation energy (S$_1 \leftarrow$ S$_0$) in the $N$ = 5, 9, 14, and 18 paraphenylenes. The coordinate and momentum, $\left|\left(P_1\right)_{mn}\right|$, density matrices for all of the other paraphenylenes can be found in the Supplementary Information. In Figure 6, the $x$- and $y$-axes represent the benzene repeat units in the molecule, and the individual matrix elements are depicted by the various colors.

Although the cyclic and linear paraphenylenes are composed of similar benzene repeat units, the density-matrix delocalization patterns in each system are considerably different. For the linear systems, the electron-hole pair created upon optical excitation becomes primarily localized in the middle of the molecule and away from the edges. In contrast, the density matrices for the cyclic systems have significant off-diagonal elements which persist even for the largest $N$ = 18 nanoring. The magnitude of the off-diagonal elements represents electronic coherence between different atoms, and Figures 6 shows that the electron-hole states are delocalized over the entire circumference in the cyclic systems.

The coherence size, which characterizes the distance between an electron and a hole, is given by the width of the momentum density matrix along the coordinate axes in Figure SI-2 (Supplementary Information). In this work, I arbitrarily define the coherence width as the distance where the momentum decreases to 10% of its maximum value. These figures show that, for a given number of benzene rings, the coherence size in the linear paraphenylenes is slightly larger than their cyclic counterparts. For the smaller paraphenylenes ($N$ < 9), the coherence size in the linear systems is larger by approximately one repeat unit in comparison to the cyclic systems. As the number of benzene units increases to 18, the coherence size for both the cyclic and linear systems approach the same value of 9 repeat units. Both $\left|\left(Q_1\right)_{mn}\right|$ and $\left|\left(P_1\right)_{mn}\right|$ also show that the linear systems only have strong optical coherences induced at their center, while the optical coherences in the cyclic geometries are nearly equally distributed throughout the entire molecule. These different density-matrix delocalization patterns and coherence sizes result in distinct excitonic properties in the cyclic and linear systems, an effect which I quantify in the next section.

### 3.3. Excitonic Effects

In order to understand electron-hole interactions on a more quantitative level, I calculated exciton binding energies for the cyclic and linear paraphenylenes as a function of size. As illustrated in Figure 7, the exciton binding energy, $E_{exc}$, is given by the difference between the quasiparticle energy gap [ionization potential (IP) – electron affinity (EA)] and the optical absorption gap ($E_{opt}$). Both the IP and EA were obtained from PBE0/6-31+G(d,p) electronic energy calculations on the $N-1$, $N$, and $N+1$ electron systems at the neutral PBE0/6-31G(d,p) optimized geometry. In this work, the vertical IP is defined by

$$IP = E_{N-1} - E_N,$$  (3)

and the magnitude of the vertical EA is given by

$$EA = E_N - E_{N+1}.$$  (4)

With these definitions, the magnitude of the exciton binding energy is then

$$E_{exc} = IP - EA - E_{opt}.$$  (5)

Table 2 gives ionization potentials, electron affinities, and exciton binding energies for the cyclic and linear paraphenylenes, and Figure 8 shows the quasiparticle energy gap (IP – EA) as a function of size.

As Figure 8 shows, with the exception of some small fluctuations, the quasiparticle energy gap in both systems decreases with size, as expected from quantum-confinement effects. Furthermore, the quasiparticle energy gap for the linear paraphenylenes decreases much more rapidly than the energy gap for the corresponding cyclic systems. The question therefore arises: Why does the quasiparticle energy gap behave so differently in the linear and cyclic systems? To address this question, we must take a closer look at the variations in aromatic character within the linear and cyclic geometries. First, in each of these organic systems (regardless of molecular geometry), an electronic competition exists between maintaining the aromaticity of the individual benzene rings versus delocalization along the backbone chain. In the unstrained linear paraphenylenes, there is relatively little conjugation along the backbone of the system since the $\pi$ electrons are localized in each individual phenyl unit to maintain aromaticity. In contrast, the cyclic paraphenylenes have highly-strained geometries which distort the electronic structure of the individual phenyl units. The strong deformation within the phenyl ring diminishes the overlap of $\pi$ orbitals, resulting in quinoidal character (antibonding interactions within the phenyl ring and double-bond character connecting adjacent phenyl rings). As a result, the electronic states in the cyclic paraphenylenes are more delocalized in comparison to their acyclic counterparts, in agreement with the transition density matrix analysis discussed in Section 3.2. More importantly, the quinoid form is energetically less stable than the aromatic form: the quinoid structure has a smaller quasiparticle gap because it involves destruction of aromaticity and a loss in stabilization energy. In other words, by increasing the quinoid character of the system, the highest-occupied orbital in the cyclic paraphenylenes becomes destabilized (i.e., raised in energy), and the quasiparticle energy gap is significantly reduced.

To give a quantitative measure of aromaticity in these systems, I calculated the nucleus-independent chemical shift (NICS(1)) at the PBE0/6-31G(d,p) level of theory for each of the phenyl rings in the cyclic and linear geometries. In the NICS(1) procedure suggested by Schleyer et al.,[33] the absolute magnetic shielding is computed at 1 Å above and 1 Å below the phenyl ring center. The resulting average NICS(1) values give a measure of $\pi$-orbital aromaticity, with more negative NICS(1) values denoting aromaticity, and more positive values corresponding to quinoidal character. As a reference point in this work, the NICS(1) value for benzene at the PBE0/6-31G(d,p) level of theory is -11.5 ppm. Table 2 gives average NICS(1) values of phenyl rings in the cyclic and linear paraphenylenes, and Figure 9 shows the NICS(1) values as a function of size. As anticipated from our qualitative discussion on aromaticity, the NICS(1) values for the cyclic paraphenylenes are always less negative (i.e., less aromatic) than their acyclic counterparts. In particular, for the smaller nanorings ($N < 8$), Figure 9 shows that these structures have significant quinoidal character, resulting in unusually small quasiparticle energy gaps (cf. Figure 8). However, as the size of the cyclic paraphenylene increases, the strain energy becomes smaller, and both the cyclic and linear NICS(1) values approach the same limit. It is also interesting to note that the fluctuations in nanoring NICS(1) values shown in Figure 9 are correlated with the deviations seen in Figures 4 and 8. Specifically, the discontinuous change in NICS(1) values at $N = 8$, 10, 13, and 17 can also be found as discontinuities in optical excitation energies (Figure 4) and abrupt changes in quasiparticle energy gaps in Figure 8. Even more intriguing is the direct correlation between NICS(1) values of individual phenyl rings and the intensity of the transition density matrix elements discussed in Section 3.2. Returning to Figure 6, the excitonic density along the diagonal for $N = 18$ is not uniform; in other words, there are 4 areas along the diagonal with maximal density at repeat units of 3, 8, 12, and 17. Figures SI-3(a) and SI-3(b) (Supplementary Information), display the NICS(1) values of individual phenyl rings and the dihedral angle between adjacent phenyl rings for the $N = 18$ cyclic geometry. Figure SI-3(a) shows 4 distinct maxima (i.e., 4 phenyl units which have quinoidal, or delocalized, character) at the same positions corresponding to maxima in the $N = 18$ transition density matrix. In contrast, Figure SI-3(b) shows that the dihedral angles are nearly identical (all within 0.3°) throughout the $N = 18$ nanoring. Taken together, these figures show that the distribution of excitonic density in Figure 6 is purely an electronic effect and not a result of conformational variations.

Finally, Figure 10 displays the exciton binding energies of both systems calculated from eq 5. In contrast to Figures 4 and 8, the exciton binding energies in both systems show a smooth monotonic variation as a function of size. As expected, for a given number of benzene rings, the exciton binding energies in the cyclic paraphenylenes are always larger (and decrease at a faster rate) than their acyclic counterparts. This trend is in agreement with the coherence sizes of the momentum density matrices discussed in Section 3.2. Compared to the linear paraphenylenes, the average electron-hole distance in the cyclic systems is smaller, leading to an increase in the binding energy. However, as the number of benzene units increases, the coherence sizes in both systems become nearly equal, and the binding energies asymptotically approach the same value. These results, in combination with the quasiparticle energy gaps and electron-hole delocalization patterns, explain the anomalous absorption energy trends in the cyclic paraphenylenes. In the linear systems, the quasiparticle energy gap (IP − EA) decreases much more rapidly than the exciton binding energy, $E_{exc}$. As a result, the behavior of the IP − EA term dominates the right-hand-side of eq 5, and the optical absorption gap ($E_{opt} = $ IP − EA − $E_{exc}$) in the linear paraphenylenes decreases as a function of size. In contrast, IP − EA in the cyclic systems decreases at a significantly slower rate than $E_{\text{exc}}$ (cf. Figures 8 and 10). Consequently, $E_{\text{opt}}$ in the cyclic paraphenylenes will be largely determined by the behavior of $E_{\text{exc}}$. Since $E_{\text{exc}}$ decreases faster than IP − EA in the carbon nanorings, the resulting optical absorption gap in the expression $E_{\text{opt}} = $ IP − EA − $E_{\text{exc}}$ *increases* as a function of size, in agreement with the TDDFT calculations.

**Conclusion**

In this study, I investigated the optoelectronic properties in a series of cycloparaphenylenes which form the molecular structure of carbon nanorings. In order to understand their unusual electronic properties as a function of molecular size, I analyzed TDDFT excited-state energies, examined electron-hole transition density matrices, and calculated quasiparticle and exciton binding energies for both the cyclic and linear paraphenylenes systems. The TDDFT calculations and transition density analyses show that the cyclic and linear paraphenylenes exhibit dramatically different size-dependent effects which can be understood in the framework of a simple exciton theory. Specifically, the main differences between the absorption spectra of the cyclic and linear paraphenylenes are caused by the subtle interplay of contributions to the electron-hole interaction energy. In the linear systems, the exciton binding energy decreases slower than the quasiparticle energy gap, and the resulting optical absorption gap decreases with molecular size, as expected. In contrast, the effective Coulomb attraction between an electron-hole pair is larger in the cyclic geometry and decreases at a faster rate than the quasiparticle band gap. As a result, the overall optical absorption gap in the cyclic paraphenylenes increases as a function of size.

In conclusion, I have shown that the different excitonic effects in the cyclic and linear paraphenylenes play a vital role in determining their distinct photophysical properties. The unique spatial distributions of electron-hole states, as depicted by their transition density delocalization patterns, result in different electron-hole interaction energies which ultimately control the electronic excitations in these systems. Looking forward, it would be very interesting to see if the unusual optical trends resulting from the lowest, weakly-allowed transition in these nanorings can be observed spectroscopically. The most "direct" way to observe these lowest transitions would be via low-temperature spectroscopic studies with an applied magnetic field to brighten the dark excitons. A magnetic flux through the open center of the nanorings would distort the circular symmetry of the electronic wavefunction and produce bright excitons due to the Aharonov-Bohm effect.[34,35] It is also possible that one can already observe these lowest transitions spectroscopically without the use of an applied magnetic field. Since these nanorings are significantly more flexible than carbon nanotubes, the lowest dark exciton may be strongly allowed through symmetry breaking. Vibronic coupling, which allows a borrowing of intensity from an energetically-close bright state, may also play a significant role in observing these weakly-allowed transitions.[36,37] For the larger carbon nanorings in particular, the first bright state (i.e. the $S_2$ state) is energetically close to the lowest dark state. As a result, the anomalous size scaling of optical excitations should be more easily observed in the larger carbon nanorings. From a theoretical point of view, it would also be interesting to see how these effects change (or if they change at all) when the paraphenylenes are functionalized with both electron donor and acceptor functional groups to facilitate charge-transfer. Calculations of this type would require the use of recent long-range-corrected density functional methods,[38-40] which are currently underway in my group.[41-43] With these goals in mind, I anticipate that a critical understanding of excitonic effects in molecular nanostructures will provide a step towards their implementation in optoelectronic applications.

**Acknowledgement.** This research was supported in part by the National Science Foundation through TeraGrid resources (Grant No. TG-CHE080076N) provided by the National Center for Supercomputing Applications. Funding for this effort was provided by the Readiness in Technical Base and Facilities (RTBF) program at Sandia National Laboratories, a multiprogram laboratory operated by Sandia Corporation, a Lockheed Martin Company, for the United States Department of Energy's National Nuclear Security Administration under contract DE-AC04-94AL85000.

**Supporting Information Available:** TDDFT $S_n \leftarrow S_0$ excitation energies and oscillator strengths up to $S_4$ for the cyclic and linear paraphenylenes, coordinate ($\mathbf{Q_1}$) and momentum ($\mathbf{P_1}$) density matrices for all the cyclic and linear paraphenylenes, NICS(1) values and dihedral angles for the $N = 18$ cycloparaphenylene, and Cartesian coordinates of all the optimized structures. This material is available free of charge via the Internet at http://pubs.acs.org.

1028.

**TABLE 1**: S$_1$ ← S$_0$ excitation energies and oscillator strengths for the cyclic and linear paraphenylenes. All properties were computed from PBE0/6-31+G(d,p) TDDFT calculations at PBE0/6-31G(d,p)-optimized geometries.

| Number of benzene rings | Cyclic Paraphenylenes | | Linear Paraphenylenes | |
|---|---|---|---|---|
| | $E_{opt}$ (eV) | Oscillator Strength | $E_{opt}$ (eV) | Oscillator Strength |
| 5 | 2.08 | 0.0022 | 3.91 | 1.8285 |
| 6 | 2.29 | 0.0000 | 3.78 | 2.2541 |
| 7 | 2.50 | 0.0016 | 3.71 | 2.6725 |
| 8 | 2.96 | 0.0000 | 3.66 | 3.0789 |
| 9 | 3.00 | 0.0170 | 3.61 | 3.5264 |
| 10 | 2.98 | 0.0000 | 3.58 | 3.9501 |
| 11 | 3.16 | 0.0266 | 3.54 | 4.4077 |
| 12 | 3.24 | 0.0000 | 3.54 | 4.8329 |
| 13 | 3.20 | 0.0012 | 3.52 | 5.2740 |
| 14 | 3.30 | 0.0000 | 3.51 | 5.7074 |
| 15 | 3.29 | 0.0416 | 3.50 | 6.1498 |
| 16 | 3.30 | 0.0476 | 3.49 | 6.6009 |
| 17 | 3.30 | 0.0001 | 3.48 | 7.0545 |
| 18 | 3.35 | 0.0000 | 3.48 | 7.5154 |

**TABLE 2**: Ionization potentials (IP), electron affinities (EA), exciton binding energies ($E_{\text{exc}}$), and average nucleus independent chemical shifts (<NICS(1)>) for the cyclic and linear paraphenylenes. The NICS(1) values were calculated at the PBE0/6-31G(d,p) level of theory, and all other properties were computed from PBE0/6-31+G(d,p) electronic energies on the $N-1$, $N$, and $N+1$ electron systems at the neutral PBE0/6-31G(d,p)-optimized geometries.

| Number of benzene rings | Cyclic Paraphenylenes | | | | Linear Paraphenylenes | | | |
|---|---|---|---|---|---|---|---|---|
| | IP (eV) | EA (eV) | $E_{\text{exc}}$ (eV) | <NICS(1)> (ppm) | IP (eV) | EA (eV) | $E_{\text{exc}}$ (eV) | <NICS(1)> (ppm) |
| 5 | 6.34 | 1.17 | 3.09 | -8.14 | 7.05 | 0.51 | 2.62 | -10.39 |
| 6 | 6.34 | 1.22 | 2.82 | -8.92 | 6.90 | 0.66 | 2.46 | -10.32 |
| 7 | 6.35 | 1.25 | 2.60 | -9.46 | 6.80 | 0.76 | 2.32 | -10.30 |
| 8 | 6.50 | 1.10 | 2.44 | -10.17 | 6.71 | 0.84 | 2.21 | -10.27 |
| 9 | 6.45 | 1.16 | 2.29 | -10.12 | 6.64 | 0.92 | 2.10 | -10.24 |
| 10 | 6.39 | 1.25 | 2.16 | -9.99 | 6.58 | 0.98 | 2.02 | -10.21 |
| 11 | 6.43 | 1.20 | 2.07 | -10.13 | 6.52 | 1.04 | 1.93 | -10.18 |
| 12 | 6.43 | 1.20 | 1.99 | -10.17 | 6.48 | 1.08 | 1.87 | -10.19 |
| 13 | 6.37 | 1.27 | 1.90 | -10.06 | 6.45 | 1.12 | 1.81 | -10.16 |
| 14 | 6.39 | 1.25 | 1.85 | -10.15 | 6.41 | 1.16 | 1.75 | -10.16 |
| 15 | 6.36 | 1.28 | 1.79 | -10.12 | 6.38 | 1.19 | 1.70 | -10.15 |
| 16 | 6.34 | 1.31 | 1.73 | -10.10 | 6.35 | 1.22 | 1.65 | -10.14 |
| 17 | 6.31 | 1.33 | 1.68 | -10.07 | 6.33 | 1.24 | 1.61 | -10.14 |
| 18 | 6.32 | 1.33 | 1.65 | -10.11 | 6.31 | 1.26 | 1.57 | -10.14 |

**Figure captions**

**Figure 1.** [*N*]-cycloparaphenylenes with *N* = 9, 12, and 18. Each cycloparaphenylene is composed solely of phenyl rings sequentially connected in the *para* position to form a single nanoring.

**Figure 2.** Molecular structures and atom labels for (a) cyclic paraphenylenes and (b) linear paraphenylenes. The specific atom numbers depicted in these figures define an ordered basis for generating the density matrices discussed in Section 2.

**Figure 3.** Average dihedral angle between adjacent benzene rings as a function of paraphenylene size. As the number of benzene rings increases, the average dihedral angle for the cyclic paraphenylenes asymptotically approaches the acyclic dihedral angle of 37°.

**Figure 4.** Optical excitation energies ($E_{opt}$) for the cyclic and acyclic paraphenylenes as a function of the number of benzene rings. $E_{opt}$ for the cyclic systems increases with size, while $E_{opt}$ for the acyclic systems decreases. All excitation energies were obtained from PBE0/6-31+G(d,p) TDDFT calculations at PBE0/6-31G(d,p)-optimized geometries.

**Figure 5.** Arrangements of individual transition dipole moments for (a) cyclic and (b) linear paraphenylenes during the $S_1 \leftarrow S_0$ excitation. The transition moments in the cyclic geometry effectively cancel, but the total transition moment increases proportionally with length in the linear systems.

**Figure 6.** Contour plots of coordinate density matrices (**$Q_1$**) for the lowest excitation energy ($S_1 \leftarrow S_0$) in the cyclic and linear paraphenylenes. The *x*- and *y*-axis labels represent the number of benzene repeat units in the molecule. The elements of the coordinate matrix, $Q_{mn}$, give a measure of exciton delocalization between sites *m* (*x*-axis) and *n* (*y*-axis). The color scale is given at the bottom.

**Figure 7.** Schematic representation of energy levels defining the exciton binding energy ($E_{exc}$), optical absorption gap ($E_{opt}$), ionization potential (IP), and electron affinity (EA). The exciton binding energy is given by $E_{exc} = \text{IP} - \text{EA} - E_{opt}$.

**Figure 8.** Quasiparticle energy gaps (IP − EA) for the cyclic and acyclic paraphenylenes as a function of the number of benzene rings. The quasiparticle energy gap for the acyclic paraphenylenes decreases much more rapidly than the energy gap for the corresponding cyclic systems.

**Figure 9.** Average NICS(1) values of phenyl rings in the cyclic and acyclic paraphenylenes as a function of size. More negative NICS(1) values correspond to an enhanced aromatic character of the system. All NICS(1) calculations were obtained at the PBE0/6-31G(d,p) level of theory.

**Figure 10.** Exciton binding energies (IP − EA − $E_{opt}$) for the cyclic and acyclic paraphenylenes as a function of the number of benzene rings. For a given number of benzene rings, the exciton binding energies in the cyclic paraphenylenes are larger than their acyclic counterparts.

*N* = 9

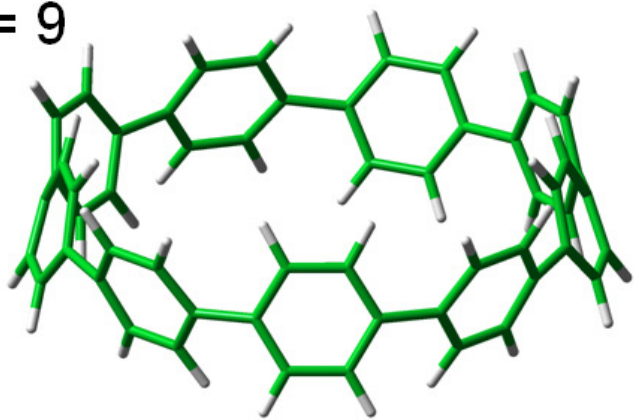

*N* = 12

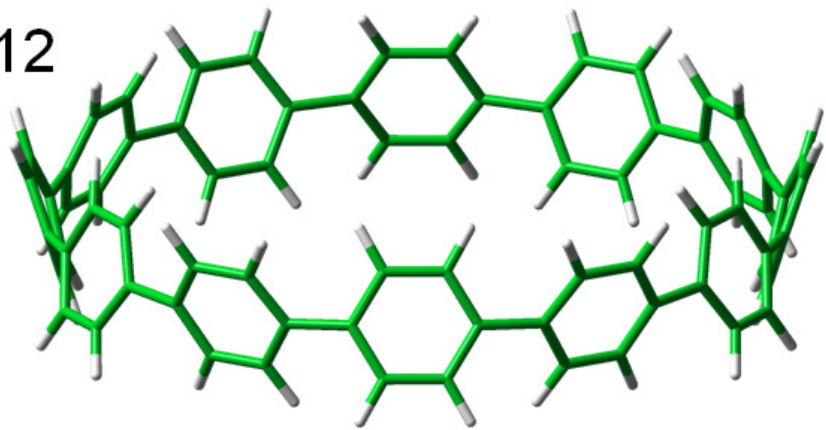

*N* = 18

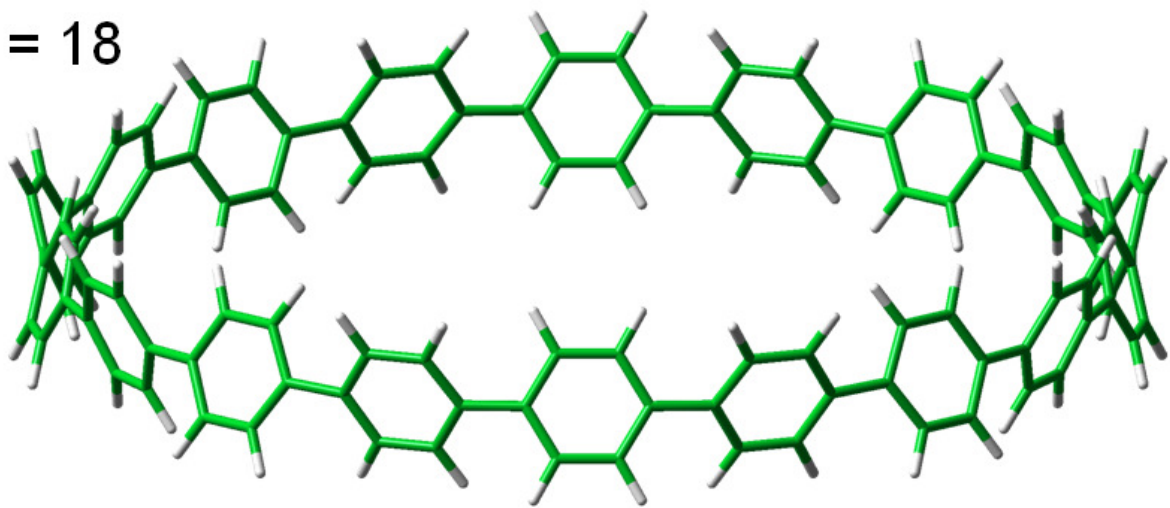

**Figure 1**

(a)

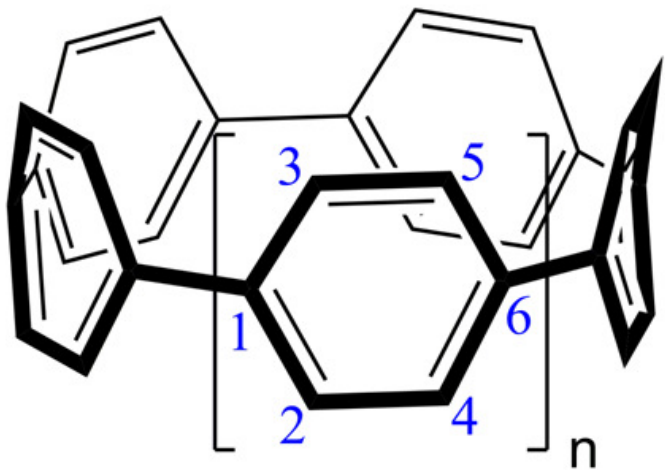

(b)

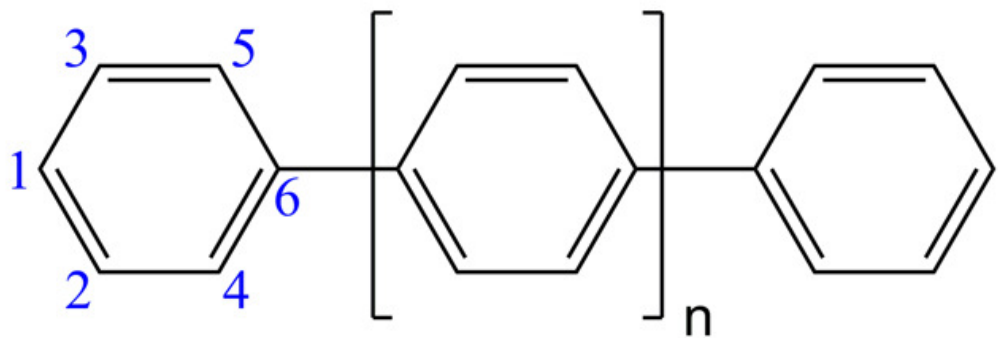

**Figure 2**

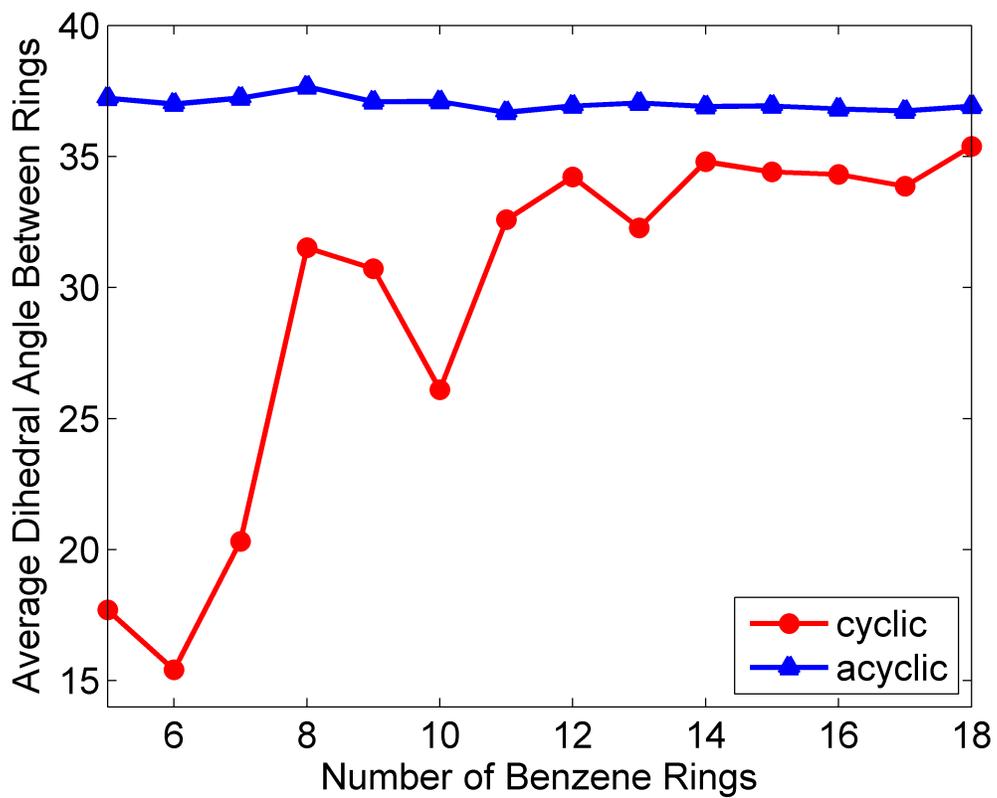

**Figure 3**

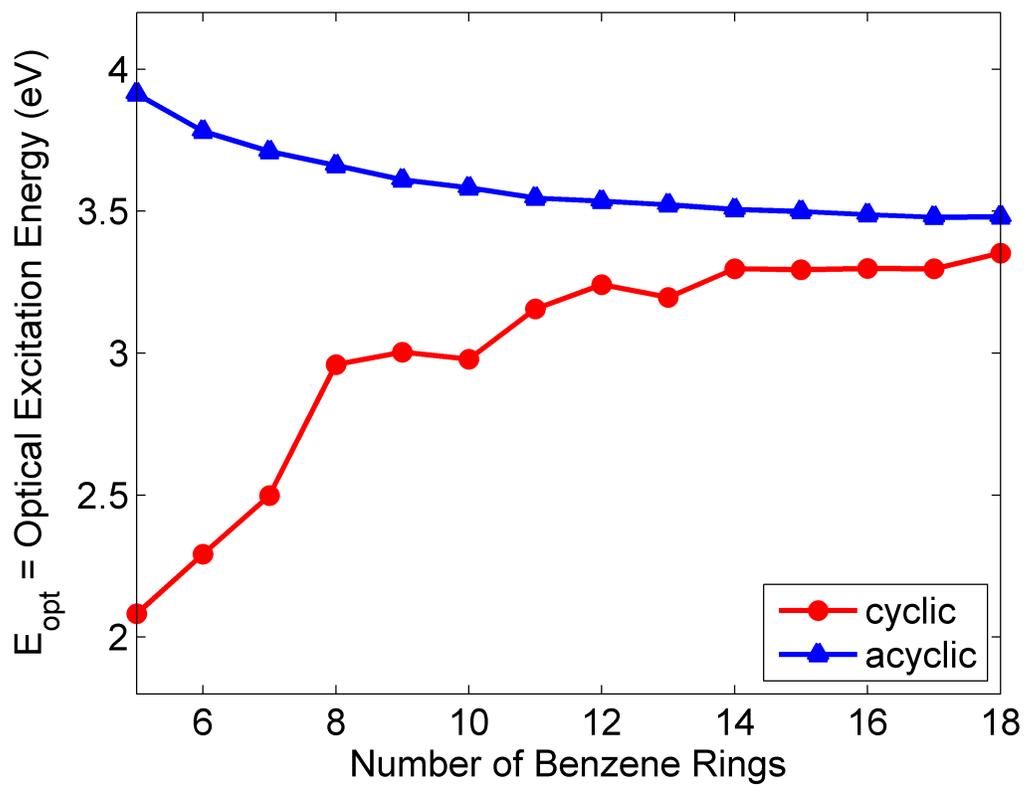

**Figure 4**

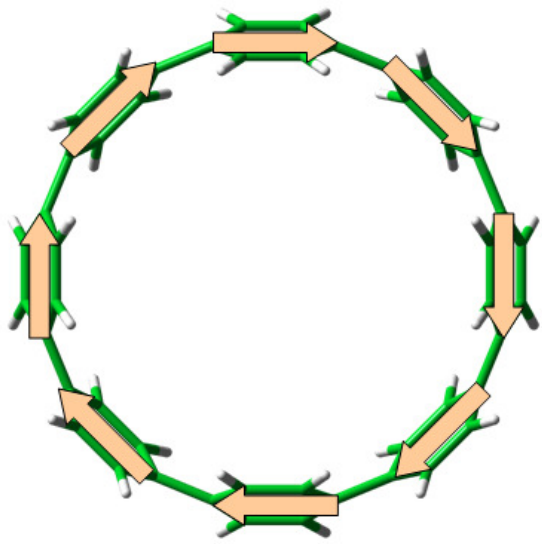

(a)

(b)

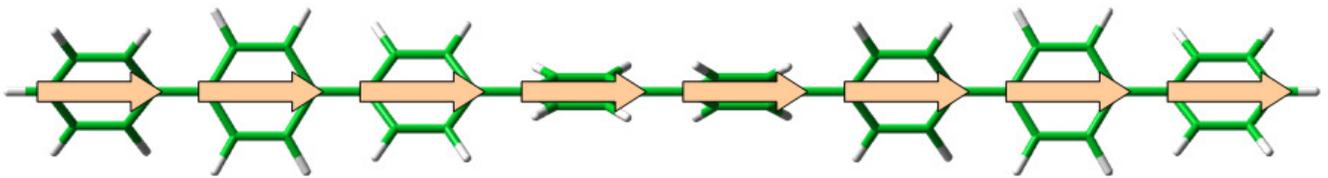

**Figure 5**

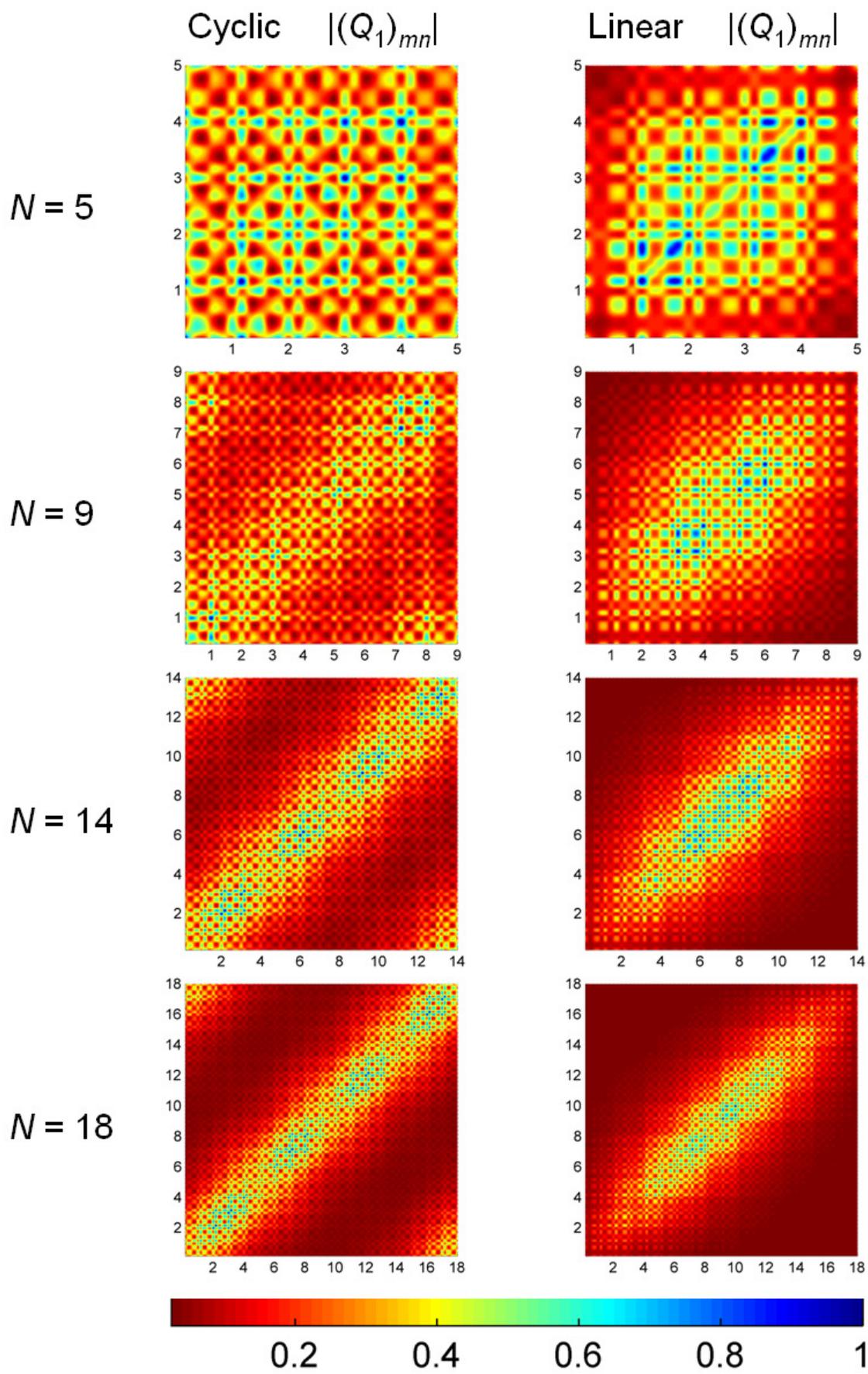

**Figure 6**

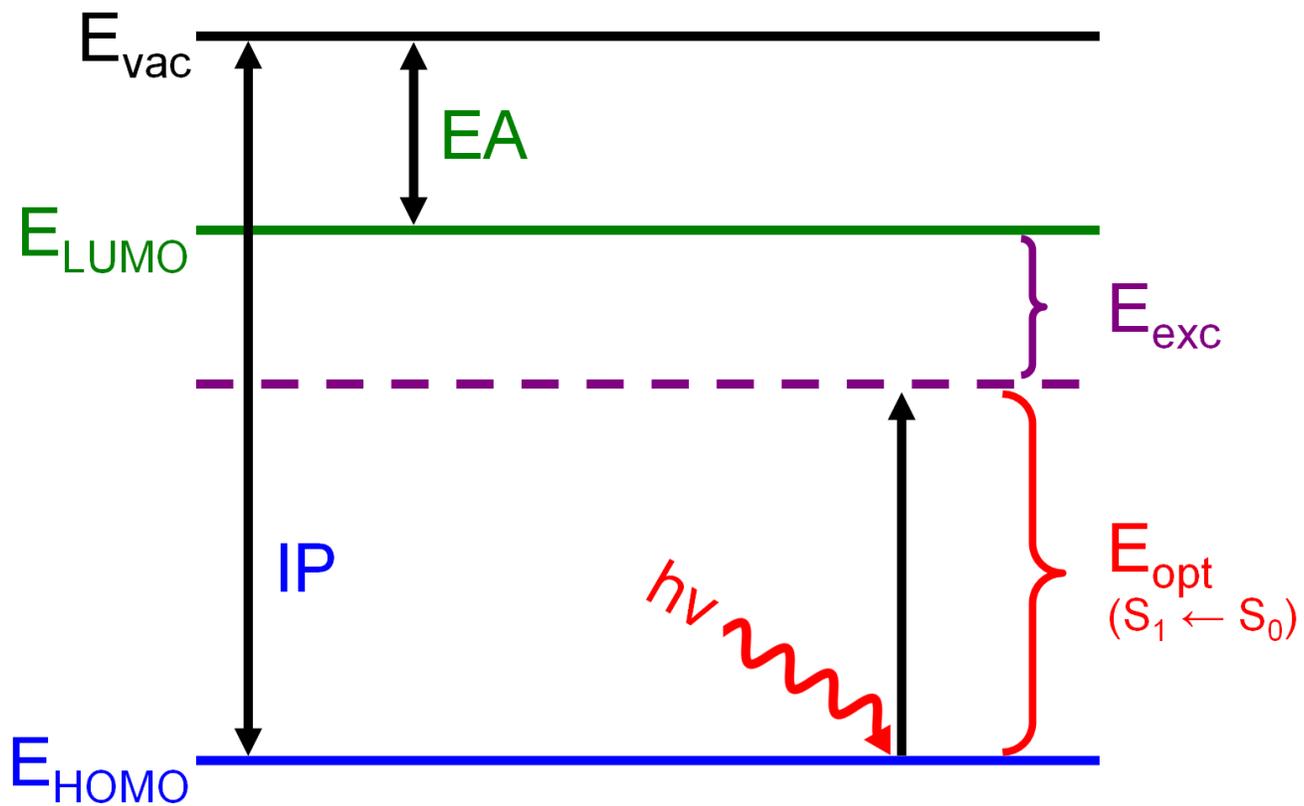

**Figure 7**

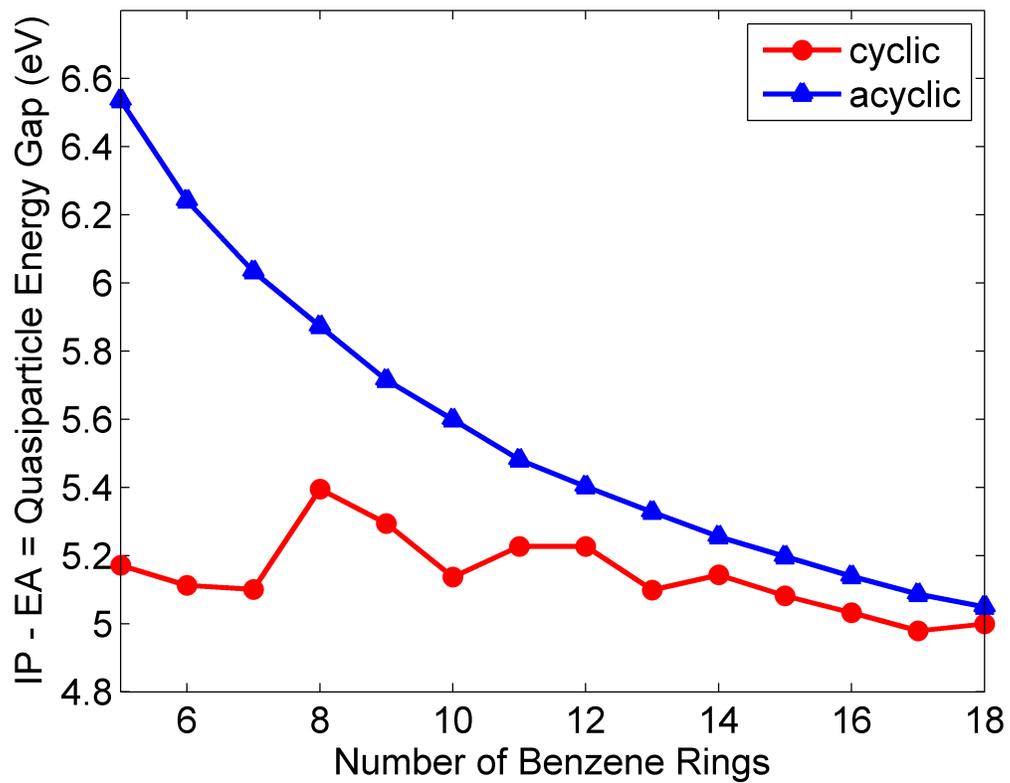

**Figure 8**

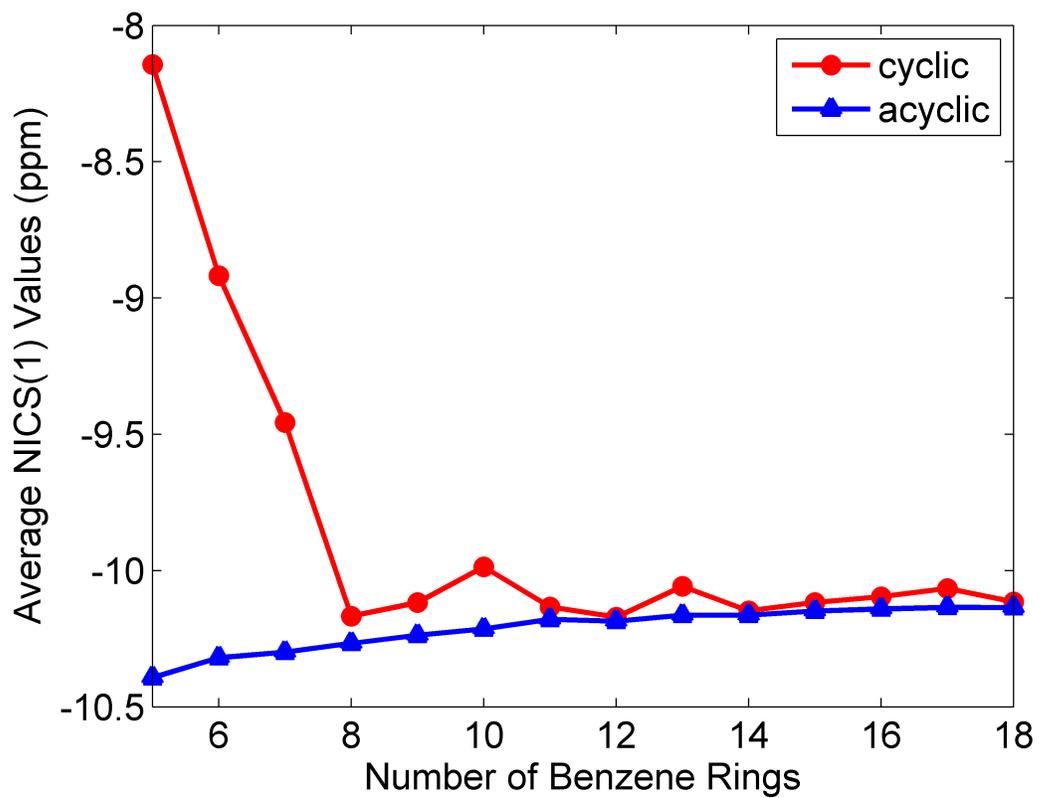

**Figure 9**

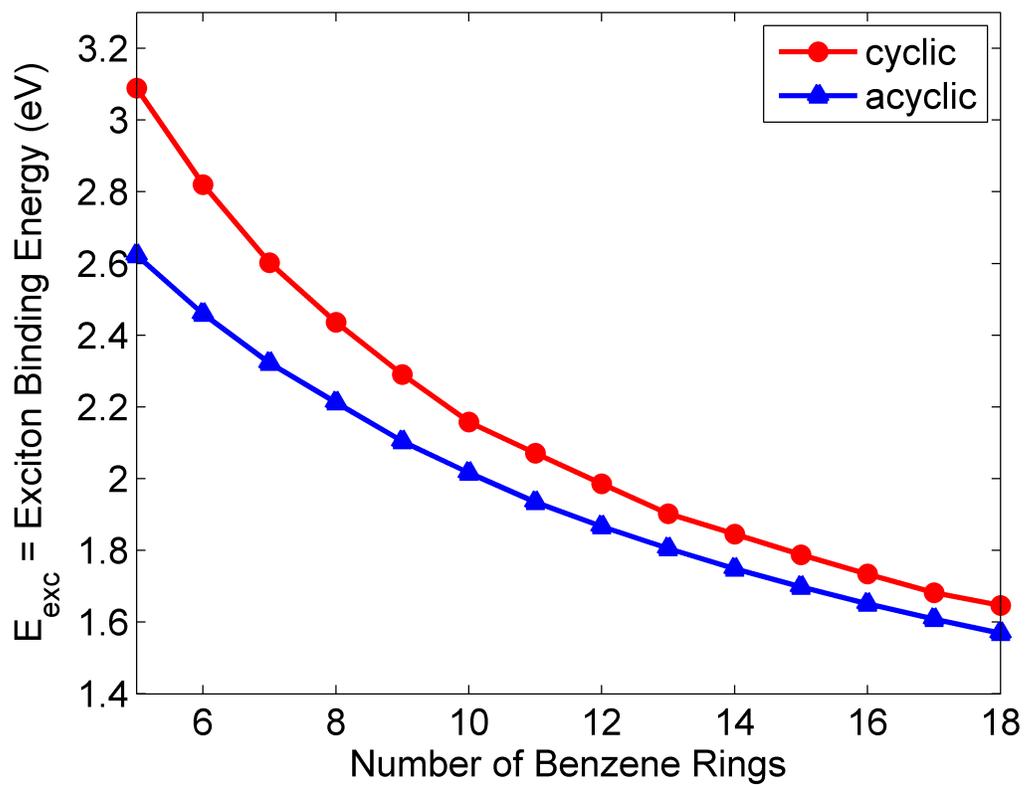

**Figure 10**